**Ultrafast light-induced formation of a metastable hidden state in bismuth vanadate**

Viktoria F. Kunzelmann,[1,2] Verena Streibel,[1,2] Philip Schwinghammer,[2] Philipp Kollenz,[3] Burak Guzelturk,[4] Franziska S. Hegner,[1,2] Lissa Eyre,[2] Frederico P. Delgado,[2] Tsedenia A. Zewdie,[1,2] Markus W. Heindl,[3] Danyellen D. Monteiro Galindo,[3] Daniel Sandner,[2] Guanda Zhou,[1,2] Elise Sirotti,[1,2] Stanislav Bodnar,[3] Yifeng Jiang,[5] Yohei Uemura,[5] Tobias Eklund,[5,6] Frederico Lima,[5] Xinchao Huang,[5] Doriana Vinci,[5] Fernando Ardana Lamas,[5] Peter Zalden,[5] Hristo Iglev,[2] David A. Egger,[2,7,*] Felix Deschler,[3,*] Ian D. Sharp[1,2,*]

[1] Walter Schottky Institute, Technical University of Munich, 85748 Garching, Germany

[2] Physics Department, TUM School of Natural Sciences, Technical University of Munich, 85748 Garching, Germany

[3] Physikalisch-Chemisches Institut, Universität Heidelberg, 69120 Heidelberg, Germany

[4] X-ray Science Division, Argonne National Laboratory, Lemont, IL, 60527 USA

[5] European XFEL, 22869 Schenefeld, Germany

[6] Institute of Physics, Johannes Gutenberg University Mainz, 55128 Mainz, Germany

[7] Atomistic Modeling Center, Munich Data Science Institute, Technical University of Munich, 85748 Garching, Germany

[*] Email: david.egger@tum.de, deschler@uni-heidelberg.de, sharp@wsi.tum.de

**Abstract**

Bismuth vanadate ($BiVO_4$) is a key photocatalyst for solar fuel applications, yet fundamental questions remain regarding the nature of photogenerated polaronic states and the lattice dynamics that govern its light-to-chemical pathways. Here, we use femtosecond optical pump–X-ray probe measurements to track the photoinduced electronic and structural dynamics in $BiVO_4$ across multiple length and time scales. Transient X-ray absorption spectroscopy captures sub-picosecond electron localization within $VO_4$ tetrahedra, consistent with small polaron formation, whereas time-resolved X-ray diffraction reveals a slower, multi-picosecond lattice reorganization into a hidden photoexcited state that is structurally distinct from both the monoclinic ground state and the high-temperature tetragonal phase. Supported by density functional theory, we show that hole-lattice interactions dynamically reduce the ground state monoclinic distortion, stabilizing the hidden state. Our results demonstrate that electron- and hole-lattice coupling jointly shape the excited state landscape, with implications for carrier transport, interfacial energetics, and light-to-chemical energy conversion pathways.

**Introduction**

The conversion of solar energy to synthetic fuels and chemicals offers a promising route to achieving a sustainable energy future. While photocatalytic semiconductors enable the direct conversion of sunlight into chemical energy, the excited state properties and interactions that underlie light-to-chemical transformations remain poorly understood, especially in the complex materials used to drive such reactions. As a notable example, monoclinic scheelite bismuth vanadate ($BiVO_4$) has emerged as a best-in-class oxide semiconductor for solar water splitting, serving as a key light harvesting element in many of the most advanced artificial and semi-artificial photosynthetic systems reported to date[1–8]. However, the prominent role of $BiVO_4$ across photocatalytic and photoelectrochemical (PEC) applications is surprising considering its vanishingly small electron mobilities, which arise from strong electron-phonon coupling[9–13]. The correspondingly large electron localization energy, in excess of 500 meV, leads to room temperature electron drift mobilities as small as $10^{-4}$ $cm^2$ $V^{-1}$ $s^{-1}$ [14,15] and plays a crucial role in limiting photovoltage generation[7,16]. Consistent with these observations, optical transient absorption studies have shown indications for sub-picosecond carrier localization, accompanied by long-lived photoexcited states[17,18]. These experimental observations agree with first-principles calculations, which have predicted electrons to form small polarons centered on vanadium (V) sites[9,10,19]. In contrast, several computational studies have predicted that holes form large polarons with much smaller localization energies, though a clear consensus regarding the nature of these states remains elusive[10,20–23]. Ultimately, improved understanding of photocarrier-lattice coupling, as well as the dynamic non-equilibrium physical and electronic structures of the photoexcited semiconductor, is critical for controlling the processes that govern the conversion of sunlight to stored chemical energy.

In this work, we apply a unique combination of ultrafast time-resolved X-ray absorption spectroscopy (tr-XAS) and X-ray diffraction (tr-XRD) measurements (**Fig. 1**)[24] to track photoinduced electronic and structural dynamics in $BiVO_4$. Together with complementary first-principles modeling, our measurements reveal a sequence of processes spanning multiple length and time scales following photoexcitation of $BiVO_4$, beginning with sub-picosecond electron localization and culminating in a slower, multi-picosecond transformation into a previously unknown metastable structure exhibiting reduced symmetry breaking. We identify this non-thermal photoexcited structure as a hidden state that emerges from a reduced driving force for the monoclinic distortion following depopulation of valence band (VB) orbitals. Our findings show that the photoexcited state of BiVO$_4$ involves a complex coupling between local electronic and collective structural dynamics, highlighting the conceptual limitations of static ground state pictures that are conventionally used to describe semiconductor function. This improved understanding of how structural dynamics shape the excited state landscape of BiVO$_4$ provides a mechanistic basis for identifying related materials in which analogous hidden states arise and for determining their influence on light-to-chemical energy conversion processes.

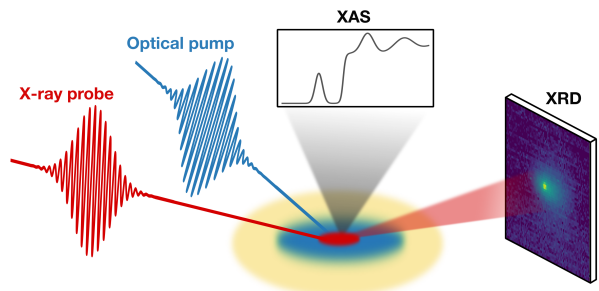

**Figure 1. Elucidating electronic and structural dynamics via combined tr-XAS and tr-XRD.** Schematic illustration of the experimental configuration for optical pump–X-ray probe measurements of the time-resolved XRD and XAS from BiVO$_4$.

**Ultrafast photoinduced electronic state dynamics**

To probe ultrafast photocarrier localization at femtosecond timescales after optical excitation, we performed tr-XAS at the V K-edge using ~50 nm thick epitaxial BiVO$_4$ (epi-BiVO$_4$) thin films on yttria-stabilized zirconia (YSZ) substrates (see Methods Section). The V K-edge, which can be measured alongside tr-XRD (see below), gives complementary information about local coordination environments surrounding V lattice sites, thus providing insights into correlated changes of the electronic and lattice structures. The ground state spectrum prior to optical pumping (**Fig. 2a**) consists of a main edge feature at ~5485.0 eV from dipole-allowed V 1s → 4p transitions, and a pre-edge at ~5471.3 eV from V 1s → 3d transitions. Although these 1s → 3d transitions are formally dipole forbidden, their strength can be enhanced in vanadium compounds by V 3d–ligand hybridization, V 3d–4p mixing, and local symmetry reductions[25–29], thereby making the pre-edge a sensitive probe of electronic structure. Here, complementary density functional theory (DFT) calculations, performed using the supercell-core-hole (SCH) method in VASP[30], reproduce the line shapes of both the main- and pre-edge of the ground state spectrum well (**Fig. 2a**). Analysis of the electronic density of states (DOS) confirms that the pre-edge corresponds to transitions into hybridized V 3d–O 2p states with minor Bi p contributions (**Fig. S1**). The asymmetric broadening towards lower energies can be attributed to crystal field splitting of the V d manifold into *e* and *t$_2$* states, with weaker transitions into *e* states arising from reduced hybridization with O 2p orbitals compared to the *t$_2$* states (**Fig.**

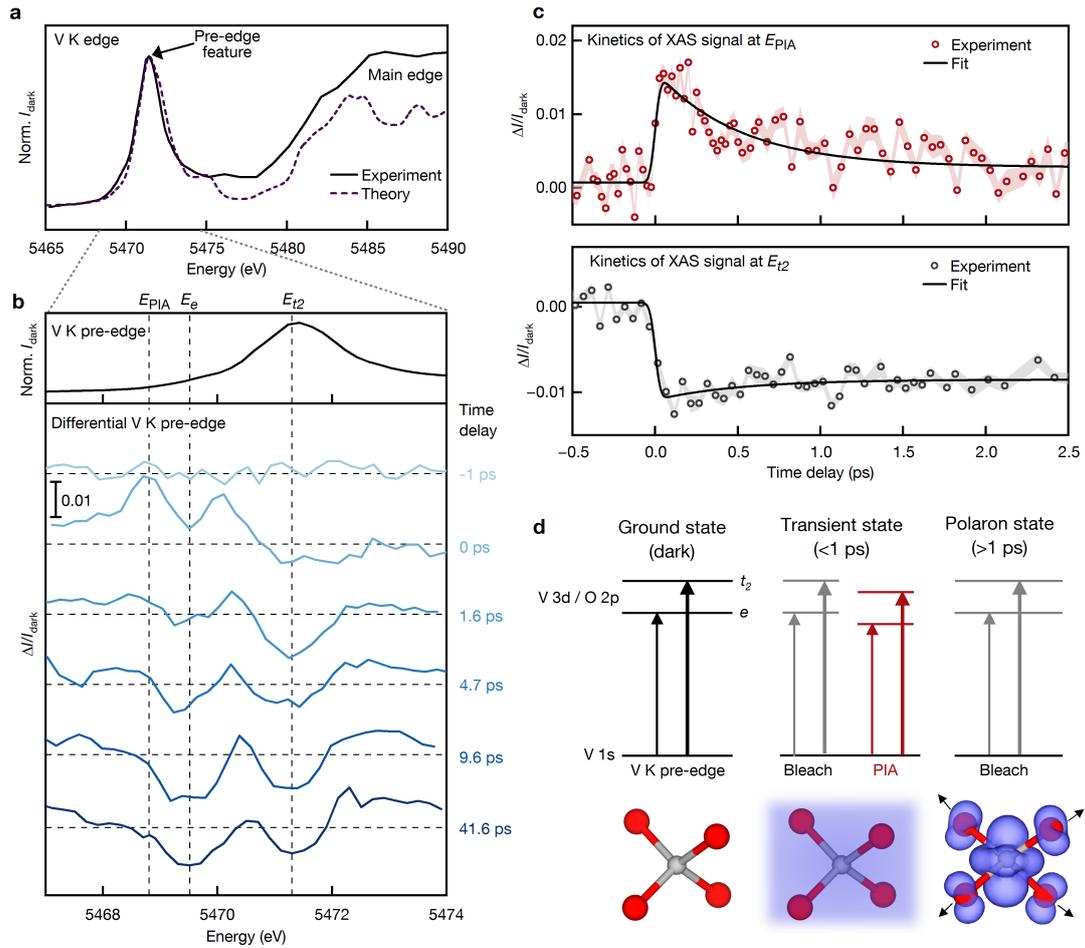

**Figure 2. Photocarrier localization dynamics probed by time-resolved XAS of epi-BiVO₄.** (a) Experimental (solid line) and simulated (dashed line) ground state V K-edge XAS spectrum of epi-BiVO$_4$, including a pronounced pre-edge feature at 5471.3 eV. (b) Differential XAS spectra collected from the pre-edge region at selected delay times following 420 nm excitation with a fluence of 2.5 mJ cm$^{-2}$. A transient photoinduced absorption feature appears near 5468.8 eV ($E_{PIA}$) and decays within the first picosecond, while bleach features at 5469.5 eV ($E_e$) and 5471.3 eV ($E_{t2}$) develop and persist to longer times. Vertical dashed lines indicate the energies of the kinetic traces in (c). (c) Temporal evolution of the $E_{PIA}$ and $E_{t2}$ features, revealing instrument response-limited FWHM rise times (~65 fs), sub-picosecond decay of the induced absorption, and a long-lived bleach. (d) Schematic representation of the V K-pre-edge XAS transitions into crystal field split $e$ and $t_2$ states. The colored arrows represent the ground state transition (black), the transient photoinduced absorption (red), and the transient bleach (grey). The lower images illustrate the corresponding evolution of the VO$_4$ units from the ground state configuration to the cases with delocalized carriers and, later, with localized small electron polarons. O atoms are shown in red, V atoms in grey.

S2). Thus, the pre-edge feature reflects hybridization of V 3d–O 2p conduction band (CB) states and local bonding in BiVO$_4$, providing us with a spectroscopic probe of electron localization and associated polaronic distortions of VO$_4$ units during tr-XAS measurements. To quantify such changes, we analyze differential X-ray fluorescence intensity, $\Delta I/I_{dark}$, defined as the difference between the pumped and unpumped XAS signal, $\Delta I$, normalized to the ground state signal of the non-photoexcited material, $I_{dark}$.

Upon above-bandgap photoexcitation with 420 nm (2.95 eV) pump pulses, the pre-edge feature undergoes pronounced transient changes (**Fig. 2b**). The differential tr-XAS spectra reveal a short-lived (< 1.6 ps) photoinduced absorption (PIA) near 5468.8 eV ($E_{PIA}$) that overlaps at the earliest times (0 ps delay) with a sharp bleach at 5469.5 eV ($E_e$). In addition, a broad bleach near 5471.3 eV ($E_{t2}$) gradually evolves into a well-defined feature at longer times (> 2 ps), resulting in a pair of bleach features at $E_e$ and $E_{t2}$ that persist to much longer delay times. The ~1.8 eV separation between these bleach features corresponds closely with the DFT-determined value of 1.6 eV for the crystal field splitting of the $e$ and $t_2$ states of $VO_4$ tetrahedra in $BiVO_4$ (**Fig. S2**) and indicates a reduced X-ray absorption strength into V 3d final states following photoexcitation.

Kinetic analysis reveals that both the PIA at $E_{PIA}$ and the bleach at $E_{t2}$ appear with instrument response-limited FWHM rise times of ~65 fs (**Fig. 2c** and **Supplemental Note 1**), consistent with ultrafast changes in electronic populations prior to any significant structural reorganization or lattice heating (see tr-XRD results below). However, stark differences in the decay kinetics of these two spectral features are observed. In particular, the PIA decays with an ~520 fs time constant (**Fig. 2c** and **Supplemental Note 1**), in agreement with recent optical transient absorption studies that report sub-picosecond electron localization in $BiVO_4$[17,18,31]. The abrupt appearance of the PIA and its subsequent decay on the timescale of electron localization suggest that it originates from free carrier screening and associated renormalization that broadens conduction band (CB) states to lower energies. As electron localization proceeds on sub-picosecond timescales, the screening strength diminishes, leading to decreased broadening and the decay of the PIA. Consistent with this interpretation, analysis of the $E_{t2}$ bleach reveals a weak transient component with the same ~520 fs time constant but of opposite sign (**Fig. 2c**), suggesting a renormalization-induced redshift of spectral weight toward the PIA region.

Beyond the small amplitude transient decay described above, the $E_{t2}$ bleach rises within the instrument response time of ~65 fs (**Fig. 2c**), and its spectral shape develops into a well-resolved bleach at the $t_2$ position within the first 1.6 ps (**Fig. 2b**). Both this feature and the lower energy $E_e$ bleach then persist to much longer times, extending well beyond the 42 ps measurement window (**Fig. 2b** and **Fig. S3**). The origin of these bleach features cannot be explained by Pauli blocking (state filling) since the $t_2$ levels lie high in the conduction band and are energetically inaccessible by the pump photons. Rather, the similar $\Delta I/I_{dark}$ magnitudes of the $e$ and $t_2$ bleach features indicate an overall reduction in the V 1s → V 3d transition strength. In addition, a slight increase in intensity between the $e$ and $t_2$ energies suggests a redistribution of the pre-edge spectral weight, consistent with a weakened ligand field and partially reduced d-manifold splitting upon polaron-induced V–O bond elongation. Such changes are consistent with decreased V 3d–O 2p orbital overlap and, to a lesser degree, with decreasing V oxidation state[25,29]. Thus, the observed spectral bleach features can be explained by small electron polaron formation at V sites, which leads to elongation of V–O bonds within $VO_4$ tetrahedra and reduction of $V^{5+}$ to $V^{4+}$ [9,10,32]. Importantly, these features emerge on timescales that are similar to the PIA relaxation, providing additional evidence that electron localization is associated with local polaronic distortions, as summarized schematically in **Fig. 2d**. While recent optical transient absorption studies have also inferred sub-picosecond electron polaron formation dynamics in $BiVO_4$ [17,18,31], the present tr-XAS measurements resolve the accompanying changes in V-centered coordination, capturing both the ultrafast screening dynamics and the emergence of polaron-induced V–O bond elongation. By probing these local, site-specific structural responses and their intrinsic timescales, tr-XAS reveals the microscopic electronic and coordination changes that define the initial excited state landscape prior to the collective lattice response investigated below.

**Photoinduced structural dynamics**

Given the strong carrier-lattice interactions in $BiVO_4$, we employed tr-XRD to determine whether electronic state changes and electron-phonon coupling following photoexcitation drive long-range lattice dynamics. Using the same epi-$BiVO_4$ sample, we first probed the momentum-resolved normalized differential scattering intensity, $\Delta I/I_{dark}$, of the (004) XRD reflection as a function of pump–probe delay (see Methods section and **Supplemental Note 2** regarding the choice of crystallographic convention according to the I2/b space group). Strikingly, differential tr-XRD rocking curves reveal an unexpected shift of the (004) Bragg peak to higher $Q$ values, as exemplified by the difference pattern collected at a delay of 6.6 ps (**Fig. 3a**), which shows increased scattering intensity on the higher-$Q$ side of the peak and decreased intensity on the lower-$Q$ side. This shift corresponds to a contraction of the *c*-axis of $BiVO_4$, indicating a long-range structural change that is opposite to that expected from thermal expansion. Analysis of the transient scattering intensity of the shifting peak at selected $Q$ values reveals that the contraction only begins after a several picosecond time delay, grows gradually, and reaches a maximum after ~10 ps (**Fig. 3b**). At longer time delays (> 20 ps), the signal inverts as the Bragg peak shifts back to lower $Q$ values, consistent with a dominant contribution from thermal expansion as heat is transferred to the lattice. We note that, although tr-XRD reveals a net lattice contraction at early times, tr-XAS indicates a local V–O bond length expansion. This apparent discrepancy can be reconciled by recognizing that photoexcitation also drives Bi–O bond contractions, as reported in Bi $L_3$-edge tr-XAS measurements of $BiVO_4$ [33,34]. Importantly, these contractions were previously observed on the ~10 ps time scale[33,34], comparable to the initial Bragg peak dynamics observed here and suggesting that $BiO_8$ subunits play a central role in the long-range structural response.

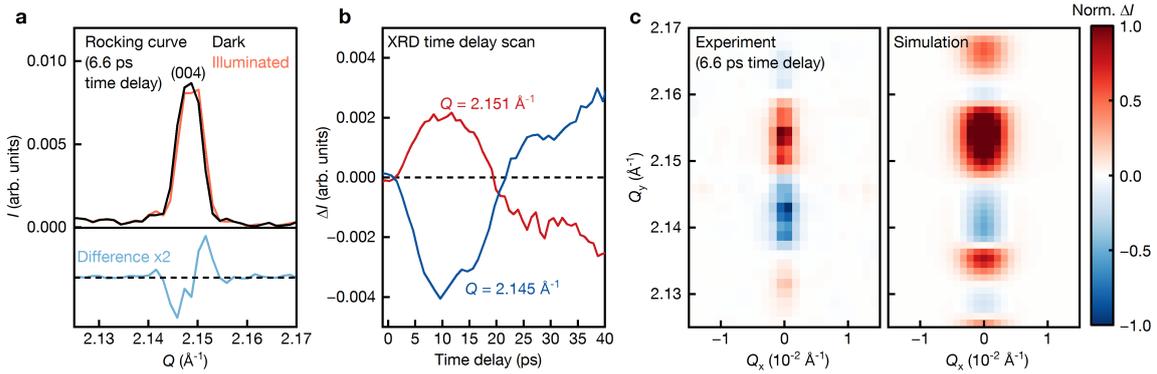

**Figure 3. Time-resolved XRD of epi-BiVO$_4$ reveals a photoinduced long-range structural transformation. (a)** Ground (dark) and excited (illuminated, 420 nm) state rocking curves of the (004) reflection at a pump–probe delay of 6.6 ps, together with the corresponding differential signal indicating a shift of the Bragg peak to higher scattering vector, $Q$. The horizonal dashed line denotes zero differential intensity. **(b)** Temporal evolution of the scattering intensities at fixed rocking curve angles corresponding to $Q$ = 2.145 Å$^{-1}$ and 2.151 Å$^{-1}$, on opposite sides of the (004) peak. The early-time decrease at $Q$ = 2.145 Å$^{-1}$ and increase at $Q$ = 2.151 Å$^{-1}$ indicate a long-range lattice contraction, whereas the subsequent reversal arises from thermal expansion at this excitation fluence (2.3 mJ cm$^{-2}$). **(c)** Diffuse scattering pattern measured at a time delay of 6.6 ps, compared with a simulated diffuse scattering pattern around the (004) reflection for a strain field extending through the complete film thickness (~44 nm).

Since shifting Bragg peak positions necessarily indicate long-range structural changes, we examined the spatial extent of the photoinduced strain. To do so, we compared simulations of diffuse scattering signals expected from modelled strain fields with measured rocking curves around the (004) reflection at a time delay of 6.6 ps (**Fig. 3c**). The experimental scattering signals are strongly localized in $Q$-space and are well-described by simulations of strain fields extending over several tens of nanometers. This finding indicates that the photoinduced distortion spans the full BiVO$_4$ thickness at the studied time delays, consistent with a quasi-homogeneous optically-induced distortion of our ~50 nm thick films. The observed *c*-axis lattice contraction occurs after several picoseconds, much slower than the sub-picosecond electron polaron localization times observed by tr-XAS, but consistent with the acoustic time scale of a propagating strain wave that establishes the distortion throughout the film.

To investigate the full crystal phase evolution, we extended the tr-XRD measurements to polycrystalline BiVO$_4$ (poly-BiVO$_4$, see Methods section) thin films, whose random grain orientations enable probing of the transient response over a broad region of momentum space with our detector geometry. At both high (**Fig. 4a**) and low (**Fig. 4b**) pump fluences (see **Supplemental Note 3**), the (112) and (004) reflections shift towards higher $Q$ on ~5-10 ps timescales after photoexcitation, consistent with the photoinduced lattice contraction observed in epi-BiVO$_4$. Furthermore, changes in reflections that are unique to the monoclinic phase indicate that these structural dynamics are associated with increased crystallographic symmetry. In particular, the intensity of the (002) reflection, which is allowed in the monoclinic but forbidden in the tetragonal phase, is reduced. Likewise, the (200)/(020) reflections, which are split in the low symmetry monoclinic phase, partially merge (**Fig. S4)**. While a small decrease in the mon-oclinic-specific (002) reflection is observed, no Debye-Waller-like changes in Bragg peak intensities appear on sub-picosecond timescales, indicating that the V–O bond elongations revealed by tr-XAS remain localized to individual unit cells. Together, these tr-XRD patterns confirm a reduced monoclinic distortion and photoinduced transition towards tetragonal symmetry on much slower timescales than the ultrafast electronic localization observed by tr-XAS.

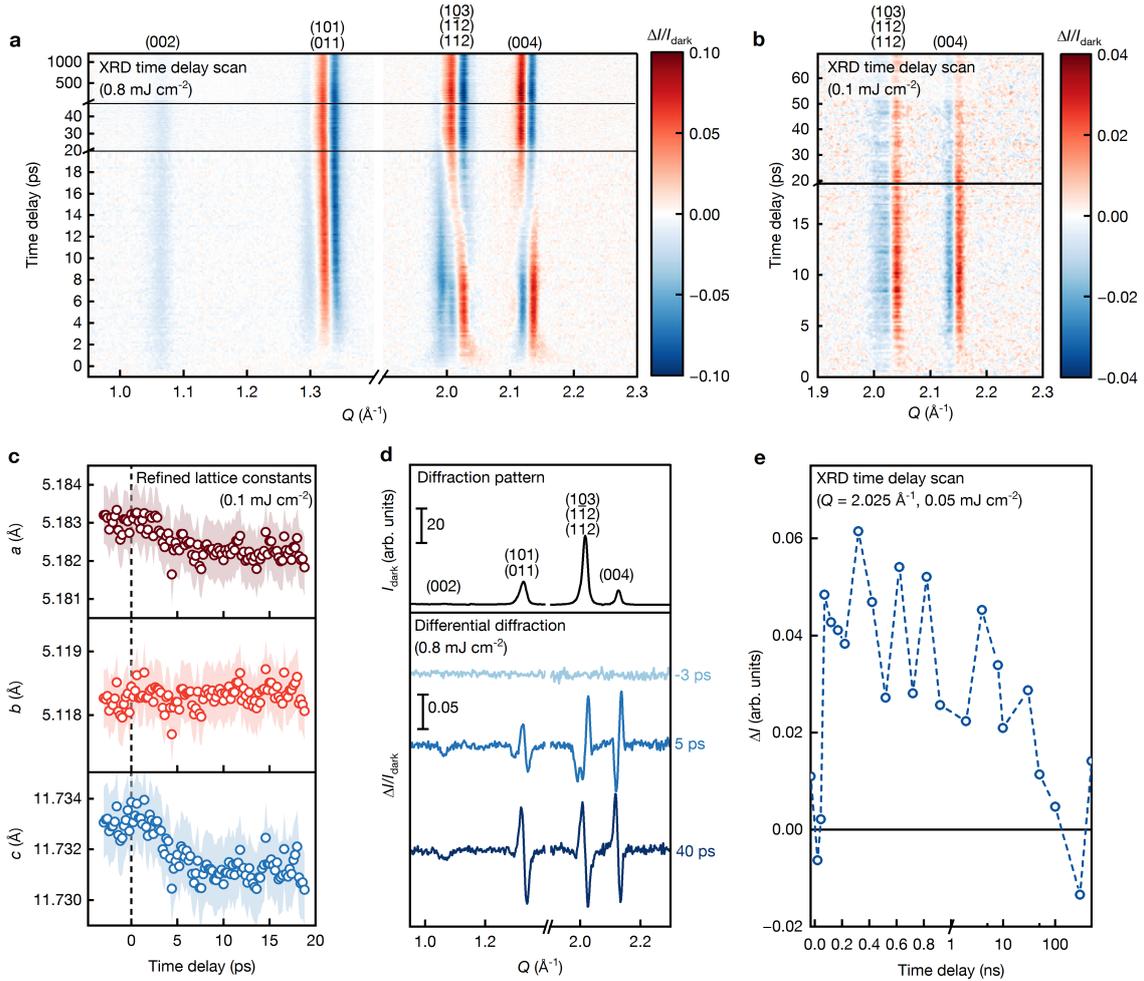

**Figure 4. Time-resolved XRD of poly-BiVO$_4$ reveals the structure of the photo-excited hidden state. (a-b)** Temporal evolution of the differential pump–probe XRD pattern from poly-BiVO$_4$ excited with a 400 nm pump beam at a fluence of **(a)** 0.8 mJ cm$^{-2}$ and **(b)** 0.1 mJ cm$^{-2}$. The early-time contraction in the (103)/(1-12)/(112) and (004) reflections indicates the formation of a photoexcited hidden state. Under high fluence excitation conditions (a), thermal expansion effects dominate at longer times but are suppressed under lower fluence excitation conditions (b), leading to the formation of a persistent photoexcited hidden state. **(c)** Time evolution of the change of the unit cell lattice parameters *a*, *b,* and *c* obtained from Rietveld refinement of the tr-XRD obtained with an excitation fluence of 0.1 mJ cm$^{-2}$. Shaded regions represent the error bars associated with structure refinement. **(d)** Differential diffraction patterns obtained at fixed delay times following excitation with 400 nm pulses at a fluence of 0.8 mJ cm$^{-2}$, highlighting the non-thermal lattice contraction at short times and thermal expansion at longer times when excited at high fluence. **(e)** Synchrotron-based tr-XRD kinetics showing that the hidden state persists with a lifetime approaching 100 ns under low fluence excitation.

Rietveld refinement of the tr-XRD patterns from poly-BiVO$_4$ provides quantitative insights into dynamic changes of the lattice constants. For low fluence excitation, where later-time thermal effects are minimized, the refinements confirm a persistent contraction along the *a* and *c*-axes (**Fig. 4c**). These anisotropic lattice dynamics demonstrate that photoexcitation drives a reduced monoclinic distortion, while decreasing the unit cell volume (**Fig. S5**).

Given the reported thermally-induced second order monoclinic-to-tetragonal phase transformation of BiVO$_4$[35], we carefully consider the role of photothermal effects. Importantly, the observed lattice contractions are distinctly non-thermal, since thermal expansion would shift the (112) and (004) reflections to lower $Q$. Indeed, at later times (>20 ps) and high pump fluences, these peaks do shift to lower $Q$ (**Fig. 4d**), marking the onset of a major contribution from thermal expansion. However, at low pump fluences, the non-thermal contraction dominates until the longest probe times (~70 ps, **Fig. 4b**). Complementary synchrotron-based tr-XRD measurements up to nanosecond time delays confirm this observation, indicating hidden state lifetimes that extend to ~100 ns under low fluence (0.05 mJ cm$^{-2}$) excitation (**Fig. 4e**), whereas thermal effects begin to dominate at earlier times at higher excitation intensities (**Fig. S6**). These results demonstrate that low-fluence optical excitation drives BiVO$_4$ into a persistent non-thermal metastable structure, thereby creating a photogenerated hidden state that is distinct from both the ground-state monoclinic and the high-temperature tetragonal structures.

Taken together, the tr-XRD measurements reveal the emergence of a photoinduced hidden state of BiVO$_4$ that is characterized by a lattice contraction and reduced monoclinic distortion, suggesting a strong coupling between photoexcited electronic states and the long-range structural degrees of freedom. This transformation develops on multi-picosecond timescales, consistent with the acoustic time required to establish a coherent long-range lattice distortion throughout the film, and persists out to the nanosecond regime characteristic of the excited state carrier lifetimes under low-fluence excitation.

**Mechanism of long-range photoinduced structural transformations**

As a basis for understanding the mechanism of hidden state formation (**Fig. 5a**), we first consider the underlying structural character of monoclinic BiVO$_4$. The monoclinic distortion arises from displacements of Bi$^{3+}$ cations from near-central positions within BiO$_8$ dodecahedra along the [001] direction[35]. Prior work has suggested that this distortion originates from differences in symmetry-allowed orbital hybridization between the two scheelite phases (**Fig. 5b**)[36]. In the tetragonal structure, the VB maximum (VBM) is dominated by antibonding Bi 6s–O 2p orbitals. In contrast, the reduced symmetry of the monoclinic phase enables Bi 6p–6s hybridization, generating bonding Bi 6p–(Bi 6s–O 2p) states near the VBM, as well as corresponding antibonding states in the CB. This symmetry-allowed hybridization reduces the antibonding character of occupied VB states, lowering their energy and stabilizing the monoclinic phase at room temperature. To quantitatively analyze such bonding and antibonding contributions to electronic hybridization, we computed the integrated projected crystal orbital Hamiltonian population (ipCOHP) using DFT (**Fig. S7**). We find that hybridization entails enhanced Bi–O bond-ing character near the VBM, increasing the Bi–O bond strength by 6.2 meV/bond for the monoclinic compared to the tetragonal phase and confirming that symmetry-allowed hybridization is a driving force for the ground state monoclinic distortion.

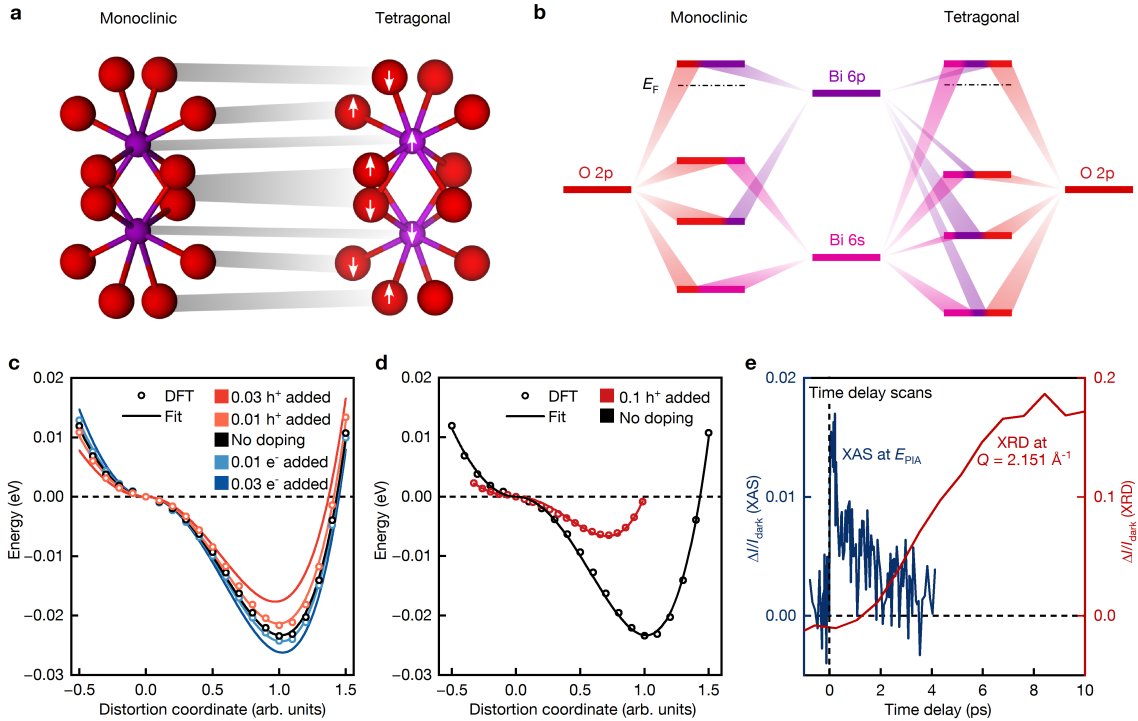

**Figure 5. Mechanism of photoexcited hidden state formation in BiVO$_4$.** **(a)** Schematic representation of the structural changes to BiO$_8$ dodecahedra between the monoclinic and tetragonal structures, with arrows indicating the atomic displacements of Bi relative to O associated with hidden state formation. O atoms are shown in red, Bi atoms in purple. **(b)** Schematic orbital hybridization diagram highlighting the symmetry-allowed Bi-O interactions that stabilize the monoclinic phase, together with the relative orbital contributions indicated by the colored bars. The Fermi level lies in the upper region of the bandgap in the investigated n-type BiVO$_4$. **(c,d)** Calculated total energies per primitive cell along the tetragonal-to-monoclinic distortion coordinate, comparing the charge neutral structure with cases containing excess carriers. Panel **(c)** shows results for low fluence conditions (0.01 and 0.03 carries per simulation cell), while panel **(d)** shows results for excess holes at a higher excitation level (0.1 holes per simulation cell, corresponding to ~2.3 mJ cm$^{-2}$). **(e)** Comparison between the ultrafast electron localization dynamics extracted from the tr-XAS decay at $E_{PIA}$, and the slower structural evolution revealed by tr-XRD of the (004) reflection (Q = 2.151 Å$^{-1}$).

To understand the influence of photoexcitation on the phase stability, we performed DFT calculations of the energetic landscape connecting the tetragonal and monoclinic phases under charge-neutral ground state conditions, as well as in the presence of either excess electrons or holes. For the charged systems, we first simulated carrier concentrations of 0.01 and 0.03 per simulation cell, comparable to the lowest experimentally generated photocarrier densities. In this low carrier-density limit, structural distortions induced by thermalized electrons and holes exclusively change the position of the minimum on the linear mode connecting the monoclinic and tetragonal structures. In this case, the photoexcited energy surface connecting the monoclinic and tetragonal structures is the average of the electron-doped and hole-doped energy surfaces, $E_{ex}(x) = (E_e(x) + E_h(x))/2$, when neglecting exciton formation[37]. **Figure 5c** shows the resulting energy landscape for each charge state as a function of the structural displacement. The energy surface is obtained by computing the total energies of crystalline geometries constructed through linear interpolation of all ionic and lattice degrees of freedom between the tetragonal and monoclinic phases. In all cases, the profiles are well described by fourth-order polynomials exhibiting the double-well potential characteristic of displacive second order phase transitions, with the monoclinic phase at the energy minimum and the tetragonal phase at the saddle point. These energy surfaces show that electrons act to stabilize the monoclinic distortion, whereas holes tend to destabilize it.

Previous theoretical work has indicated that electron doping stabilizes the monoclinic structure of BiVO$_4$[38]. While our calculations support this conclusion, they reveal that the electronic contribution to phase stability arises from two competing mechanisms that partially cancel. Electron localization on VO$_4$ tetrahedra increases the negative partial charge on oxygen, strengthening ionic Bi-O bonds. Since symmetry breaking in monoclinic BiVO$_4$ shortens a subset of the Bi-O bonds, this increase in ionic bonding favors the monoclinic phase. The polaronic elongation of V–O bonds further enhances this effect by displacing O ions towards Bi. Opposing these effects is a covalent contribution, in which the increased hybridization in the monoclinic phase raises the energy of CB states (**Figure S7**). Photoinduced electron occupation of these higher energy CB states therefore destabilizes the monoclinic structure. The resulting competition between ionic bond strengthening and covalent bond destabilization yields a net stabilization of the monoclinic phase of 0.1 eV per small electron polaron (see **Supplemental Note 4**). Thus, photogenerated electrons cannot account for the experimentally observed transition into the hidden state.

In contrast to the stabilizing influence of electrons, we find that holes decrease the energetic well depth of the energy surface (**Fig. 5c**), explaining the transition to the hidden state. This behavior arises from depopulation of Bi–O bonding orbitals near the VBM, which weakens the hybridization that stabilizes the monoclinic phase. As a result, the energy gain from Bi 6p–(Bi 6s–O 2p) mixing no longer compensates for the strain energy cost of the distortion and the lattice is driven towards the hidden state, a configuration with reduced symmetry breaking. To better understand the influence of holes at higher excitation fluences using methods compatible with small electron polaron calculations, we performed additional DFT calculations with 0.1 excess holes per unit cell, including structural relaxation, but treating the hole polarons as delocalized due to their predicted size of up to 2 nm[39]. At these carrier concentrations, the energetic well depth decreases by 8.5 meV/primitive cell and shifts the minimum towards a configuration with a reduced monoclinic distortion (**Fig. 5d**), consistent with the results in the low carrier density limit (**Fig. 5c**). By calculating the combined energetic impact of small electron polarons and thermalized holes, we find a net flattening of the double well potential energy surface by 0.07 eV per charge carrier, or 3.5 meV per primitive cell, at a concentration of 0.1 electron-hole pairs per unit cell.

While the dominant role of photogenerated holes in the long-range structural dynamics may seem surprising given the strong electron-lattice coupling observed by tr-XAS, our findings reveal a consistent connection with prior observations of carrier-phonon interactions and ultrafast dynamics in $BiVO_4$. Previous optical and THz transient absorption studies have shown that holes couple to the soft $A_g$ optical phonon mode of $BiVO_4$ (63 cm$^{-1}$), which is associated with $Bi^{3+}$ vibrations relative to $VO_4^{3-}$ tetrahedra[12,40–42]. This vibrational mode also gives rise to the Bi displacements that define the monoclinic distortion[43]. Reported hole localization times of several picoseconds[17,18,40,41] coincide with the structural dynamics observed in our tr-XRD measurements, as well as with photoinduced Bi–O bond contractions[33,34]. In contrast, electrons couple to much higher frequency V–O stretching modes ($A_g$, 820 cm$^{-1}$), favoring rapid sub-picosecond localization. Indeed, direct comparison of our tr-XAS and tr-XRD dynamics (**Fig. 5e**) reveals that electron localization occurs on sub-picosecond timescales, whereas the structural transformation associated with the hidden state only emerges after several picoseconds. Although electron polarons primarily distort $VO_4$ tetrahedra rather than the $BiO_8$ sublattice that determines the magnitude of symmetry breaking, their rapid localization reduces electronic screening and increases the partial negative charge on oxygen anions, which further enhances the influence of photogenerated holes on the Bi–O bonding environment. Overall, our computational and experimental results indicate that photogenerated holes destabilize the monoclinic phase and initiate a long-range structural rearrangement toward a nonthermal structure with reduced symmetry breaking, leading to formation of the hidden state.

**Conclusion**

By tracking the dynamic evolution of $BiVO_4$ across multiple time and length scales, we identify a sequence of electronic and structural processes that give rise to a photogenerated hidden state that is structurally distinct from both the monoclinic ground state and the high temperature tetragonal phase. These insights reveal the unique role of photoexcitation in accessing non-equilibrium regions of the lattice energy landscape, with important implications for the function of $BiVO_4$ and related oxides as light absorbers in photocatalytic and photoelectrochemical systems. In particular, the coexistence of small electron polarons with hole-induced lattice symmetry changes indicates that static ground state structural models may not capture the key photophysical processes occurring under illumination. Because these structural responses depend sensitively on the relative populations of electrons and holes, the excited state lattice is expected to evolve under illumination conditions relevant to photocatalysis, where imbalanced oxidation and reduction reaction kinetics lead to net charge accumulation[44]. Sluggish water oxidation, for example, can lead to hole accumulation that further drives the lattice toward the hidden state, thereby impacting energetics and carrier transport during operation. Thus, our findings suggest a possible mechanism by which surface reaction kinetics may influence the structural and electronic properties of active photocatalysts, mediated by carrier-lattice coupling. In addition, because hole localization in $BiVO_4$ is highly sensitive to the structural phase, the observed photoinduced reorganization is likely to significantly impact minority carrier mobilities, which are critical for efficiently driving oxidation reactions. Finally, this new understanding suggests a possible connection to the unusually long photocarrier lifetimes observed in $BiVO_4$, raising the question of whether the coexistence of localized small electron polarons, spatially extended large hole polarons, and hole-driven long-range lattice symmetry changes could introduce an additional barrier to recombination. More broadly, the identification and mechanistic understanding of this photoexcited hidden state highlights the importance of non-equilibrium responses in light harvesting semiconductors, providing new opportunities to design materials that could exploit such states for enhanced functionality.

**Methods**

**Sample synthesis.** As a basis for time-resolved X-ray studies, we prepared both epitaxial and polycrystalline monoclinic scheelite BiVO$_4$ thin films (epi-BiVO$_4$ and poly-BiVO$_4$, respectively) with thicknesses of ~50 nm using solution-based synthetic methods that have been shown to yield favorable PEC performance characteristics[45,46]. The precursor solution was obtained by mixing equal parts of 0.2 M bismuth(III) nitrate pentahydrate (Bi(NO$_3$)$_3$•5H$_2$O, Sigma-Aldrich) in acetylacetone (C$_5$H$_8$O$_2$, Sigma-Aldrich) and 0.03 M vanadium(IV)-oxy acetylacetonate (OV(C$_5$H$_7$O$_2$)$_2$, Sigma Aldrich) in acetylacetone. For a spin coating step, 500 µl of the filtered solution were spin-coated onto the substrate, followed by annealing in air at 500 °C for 10 min to decompose the metal-organic precursors. The process was repeated nine times, until the target thickness of ~50 nm was reached.

Epitaxial BiVO$_4$ films were grown on 2" YSZ (001) wafers (8 mol% yttria, undoped, Alineason) after an optimized cleaning and pre-annealing procedure, as described by Kunzelmann et al.[46] After spin coating of the final layer, the films were subjected to a final annealing sequence consisting of 500 °C for 10 min in air, natural cooling to room temperature, and a second anneal at 650 °C for 10 min in a confined air volume, which induced the transformation from initially polycrystalline to epitaxial BiVO$_4$. Polycrystalline BiVO$_4$ thin films were grown on 2" (0001) sapphire wafers that were cleaned via ultrasonication in acetone and isopropanol for 10 min each, followed by drying under flowing N$_2$. The same spin coating and intermediate annealing steps used for epitaxial films were then applied nine times to generate ~50 nm thick films. However, in this case, crystallization was achieved by a single final annealing step at 500 °C for 2 h, yielding homogeneous polycrystalline BiVO$_4$ thin films. High-resolution XRD of the BiVO$_4$/YSZ (**Fig. S8**) and grazing-incidence XRD of BiVO$_4$/sapphire (**Fig. S9**) confirmed the successful synthesis of phase-pure monoclinic scheelite BiVO$_4$ in both epitaxial and polycrystalline forms. In addition, photothermal deflection spectroscopy (**Fig. S10**) verified an optical bandgap of ~2.5 eV and negligible sub-bandgap optical absorption for both materials, consistent with high-quality thin films.

**Time-resolved XAS and XRD.** Time-resolved XAS and XRD measurements using X-ray free-electron laser radiation were conducted at the FXE instrument of the European XFEL[24]. The optical pump was provided by the integrated pump–probe laser system, which generated 800 nm fundamental pulses of 50 fs duration from a non-colinear optical parametric amplifier (NOPA) that were doubled in frequency using a beta-barium borate (BBO) crystal. The relative pump–probe timing was controlled via a motorized delay stage. The pump laser provided a pulse pattern matching the X-ray delivery sequence, with 5 Hz on/off modulation for pump–probe measurements. Synchronization between X-rays and optical laser pulses yielded an instrument response function (IRF) of ~65 fs. The combined tr-XAS and tr-XRD measurements on epi-BiVO$_4$ were performed with a pump wavelength of 420 nm and the tr-XRD measurements of poly-BiVO$_4$ were conducted with a pump wavelength of 400 nm. In all cases, the pump beam was s-polarized and the corresponding fluences are specified in the main text (see **Supplemental Note 3** for details).

Hard X-ray probe pulses of ~30 fs duration were generated by spontaneous amplification of stimulated emission (SASE) in an undulator section with tunable gaps. The X-ray pulses were monochromatized by two reflections from Si(111) crystals and focused using lenses. For signal normalization, the incident pulse energy ($I_0$) was determined from the "s" scattering intensity measured at 90° from a thin layer of Kapton.

The combined time-resolved XAS and XRD experiments on epi-BiVO$_4$ were conducted using monochromatized X-rays, with a variable focus size due to dispersion of the beryllium lens material, resulting in spot sizes of 50 and 100 µm FWHM at 5.4 keV and 5.8 keV, respectively. Bursts of 16 X-ray pulses were delivered with a repetition rate of 94 kHz, repeating with a 10 Hz rate. Each probe pulse had an energy of ~25 nJ. The incident X-ray intensity was estimated by measuring the scattered signal from a thin Kapton film placed upstream of the sample chamber. A reverse-biased PIN-photodiode (detection area 10×10 mm$^2$) was used to monitor the scattered X-ray intensity and the output was read by a fast digitizer (ADQ412, Teledyne SP Devices). A fast transimpedance amplifier (DHPCA-100, FEMTO) was used to adjust the signal level from the photodiode. XAS signals were collected in total fluorescence yield (TFY) mode from the fluorescence intensity measured at an ~90° emission angle using two 28×28 mm$^2$ reverse-biased PIN-diodes, read by an ADQ412 fast digitizer. The fast transimpedance amplifiers were employed to adjust the signal level. Diffraction signals were collected with a Jungfrau 1M (JF1M) detector operated at fixed medium or low gain with 16 memory cells for pulse-resolved detection of horizontally scattered *p*-polarized X-rays. For both tr-XAS and tr-XRD, the laser/X-ray crossing angle was 2.5°.

The diffraction-only experiments on poly-BiVO$_4$ were performed with SASE X-rays at 9.3 keV photon energy, focused to a 10 µm FWHM spot size and delivered at 10 Hz burst repetition rate with 50 nJ/pulse. The same JF1M detector described above was operated in dynamic gain mode and mounted on a robot arm to collect intensity patterns at various vertical angles ("s" scattering). For these poly-BiVO$_4$ measurements, a quasi-grazing incidence geometry was used, with an X-ray incidence angle of 3.5° with respect to the sample surface plane. The laser/X-ray crossing angle was 1.5°.

The unit cell parameters as a function of pump–probe delay time were quantified by Rietveld refinement[47] of poly-BiVO$_4$ diffraction patterns collected from both the XFEL- and synchrotron-based measurements using the software GSAS-II[48]. The experimental data were fitted and refined to an initial calculated BiVO$_4$ diffraction pattern in the I2/b space group using the non-linear least squares method.

Complementary grazing incidence tr-XRD measurements of poly-BiVO$_4$ were performed at beamline 11-ID-D of the Advanced Photon Source (APS) at Argonne National Laboratory. Such measurements provided a higher signal-to-noise ratio to enable measurement over a broader range of excitation fluences, but with a reduced time resolution compared to the XFEL measurements. Above bandgap photoexcitation was provided by a 400 nm pump laser with a tunable intensity between 10 – 3000 µJ cm$^{-2}$, operated at 3 kHz repetition rate with 100 fs pulse duration. Structural changes were probed by a high-flux monochromatic X-ray beam with ~10$^{13}$ photons s$^{-1}$ generated by two in-line undulators, optimized beamline optics, and a high-heat load double crystal monochromator. The grazing-incidence angle of 1.5° resulted in an X-ray probed area of 400 µm × 1.9 mm, which was fully encompassed by the normal-incidence pump laser spot size of 0.9 × 5 mm$^2$. The hard X-ray energy was tuned to 11.7 keV to provide a sufficient penetration depth into the crystal. The repetition rate of the X-ray probe pulses was 6.5 MHz with 153 ns spacing and a temporal FWHM of 50 to 80 ps. The diffracted X-ray intensities were recorded by a single-gated silicon detector (Pilatus 2M, Dectris Ltd.). Pump–probe delays were sampled in random order and the BiVO$_4$ sample was continuously rotated during data collection to avoid photocharging effects. The BiVO$_4$ sample was stable during the pump–probe experiments, with no degradation observed up to the highest tested pump fluences.

**Strain-dependent X-ray scattering simulations.** The photoinduced displacement field was modeled as a radially symmetric deformation centered at position $(x_c, y_c)$. For each lattice point at position $(x_i, y_i)$, we calculated the distance from the center of the strain field as:

$$r_{ij} = \sqrt{(x_i - x_c)^2 + (y_i - y_c)^2} \qquad (1)$$

The radial displacement magnitude follows a Gaussian-modulated form:

$$f(r) = -\frac{rA}{r_\varepsilon} \exp\left(-\frac{r^2}{r_\varepsilon^2}\right) \qquad (2)$$

where $A$ is the displacement amplitude and $r_\varepsilon$ is the characteristic radius of the strained region. This approach produces a displacement field that is zero at the center, reaches a maximum at approximately $r \approx \frac{r_\varepsilon}{\sqrt{2}}$ and decays exponentially for large $r$. The displacement vector is pointed towards the center of the displacement field. To compute the scattered X-ray intensity, we created a regular 2D lattice of dimensions $m \times n$ with lattice sites at integer positions (Lattice parameter 1) and each lattice position was displaced according to the model above. We up-sampled the lattice by a factor $s$, creating a scattering density distribution $\rho(x,y)$ with dimensions $ns \times ms$. For each point in the up-sampled grid at position $(x,y)$, the density contribution from each displaced lattice site $(x_i, y_j)$ was calculated as follows:

$$\rho(x,y) = \sum_{ij} G(x - x_i, y - y_j) \tag{3}$$

where $G(x,y)$ is a normalized 2D Gaussian function. The scattered intensity in reciprocal space was then obtained via 2D Fourier transform:

$$I(k_x, k_y) = |\mathcal{F}[\rho(x,y)]| \tag{4}$$

The differential scattering intensity was calculated as the difference between the lattice with the strain present ("on" state) and without it ("off" state):

$$\Delta I(k_x, k_y) = I(k_x, k_y) - I_{ref}(k_x, k_y) \tag{5}$$

Both intensity maps were smoothed with a 2D Gaussian filter before subtraction to reduce high-frequency noise.

**Computational details.** Density functional theory (DFT) calculations were performed using the projector augmented wave (PAW) method as implemented in VASP [49]. For all electronic structure calculations, the self-consistent cycle was considered converged when change in total energy was below $10^{-6}$ eV, and a Gaussian smearing with a width of 0.03 eV was applied. We performed all calculations with a conventional unit cell containing two primitive cells.

The relaxed 0 K structures of monoclinic and tetragonal BiVO$_4$ were obtained using the HSE06 hybrid functional including spin-orbit coupling (SOC)[50,51]. A 4×4×2 k-point grid was employed with PAW pseudo-potentials[52] for PBE with the Bi ($5d^{10}6s^26p^3$), V ($3p^63d^34s^2$), and O ($2s^22p^4$) valence electron configurations, and the plane-wave cutoff was converged to 550 eV. Structures were relaxed using the GDIIS algorithm, as implemented in the GADGET code[53], until the norm of the forces was smaller than 0.005 eV Å$^{-1}$.

For evaluation of Bi-O bonding interactions, additional relaxations were performed with the same parameters using the HSE functional with α = 0.4, but without SOC to enable compatibility with the LOBSTER code[54–56]. The DFT results were post-processed with the LOBSTER code to compute the integrated projected crystal Hamiltonian population (ipCOHP) for Bi-O bonds in monoclinic and tetragonal scheelite BiVO$_4$.

XAS spectra were simulated using the supercell-core-hole (SCH) method, as implemented in VASP[57], with a 3×3×3 supercell of the relaxed monoclinic structure. These calculations were performed using the PBE functional with SOC and PAW pseudopotentials[52] with the Bi ($5s^25p^65d^{10}6s^26p^3$), V ($3s^23p^63d^34s^2$), and O ($2s^22p^4$) valence electron configurations. Due to the larger super cell, a 2×2×2 k-point grid was found to be sufficiently converged at a plane-wave cutoff of 550 eV. To calculate XAS spectra, an electron from the V 1s orbital of a single vanadium was placed into the conduction band within the final state approximation. The number of evaluated Kohn-Sham bands was set to 6752 and the dielectric function was evaluated on 500001 points with a broadening of 0.0001 eV. Since the SCH method only reproduces the relative peak positions, the simulated spectrum was aligned to the experimental data by fitting the Lorentzian broadening, *x*-and *y*-offsets, and *y*-scale to the pre-edge region. Details regarding the method of associated DOS calculations can be found in **Supplemental Note 5**.

To analyze the impact of photogenerated carriers on the Born-Oppenheimer potential energy surface, we sampled the energies along a linear path connecting the 0 K monoclinic and tetragonal structures, using the same settings as for the relaxations. The ambiguity in the interpolation due to translational and rotational invariance for both structures was solved by rotating the unit cells such that the mean-square deviation (MSD) between the lattice vectors was minimized, as well as by translating the cells such that the MSD between equivalent atoms was minimized. The use of a single linear mode to represent the Born-Oppenheimer energy between the two phases was motivated by prior work by Liu et al.[58], who found that a single phonon mode can be associated with the atomic displacements of the phase transition. The resulting potential energy surface represents one half of a double well potential connecting the tetragonal scheelite structure with two symmetry-equivalent monoclinic scheelite ground state structures. We performed relaxations for the monoclinic and tetragonal structures using the same settings outlined above, but with 0.1 holes added to the simulation cell to acquire the altered transition mode. Single-point energies were calculated for several structures along the ground state and hole-doped transition modes, using the same settings as for the relaxations. The resulting potential energy surfaces were fitted with a fourth-order polynomial. To compare the change in the relative distances of the tetragonal and monoclinic minima, we scaled the length of the *x*-axis with the root-mean-square deviation (RMSD) of the atomic coordinates between the monoclinic and tetragonal ground state structures, normalized to the value for zero charge doping. The RMSD was acquired using the relative rotation and translation of the cells that minimized the RMSD between the lattice vectors and the atomic coordinates of the two cells. Additionally, we performed the same calculations without relaxation of the monoclinic and tetragonal ground state structures with added charge for 0.0001, 0.001 and 0.01 electrons/holes per unit cell. We then performed a linear fit on the well-depth and the distance between the maximum and minimum along the transition mode, to obtain a simple model for the effects of electrons and holes at very low charge densities. We used this model to calculate the changes to the potential energy surface at an added 0.01 and 0.03 electrons and holes per unit cell.

To evaluate the impact of electron polaron formation on the monoclinic and tetragonal structures, we created 2x2x1 supercells of the unit cell containing two primitive cells for each. We then added a single electron to the structures and expanded the V-O bonds around one vanadium atom, calculating the energy with single-point calculations, using the same functional and other settings as in the initial relaxation. Once a minimum in the energy surface for the bond length was found, the structure was allowed to relax completely. The total energy difference between the monoclinic and tetragonal structures with and without the small electron polaron was then determined. The charge density of the polaron was extracted by calculating the all-electron density of the highest occupied band from the VASP output, using the VaspBandUnfolding code, and visualized using VESTA[59,60].

**Acknowledgements**

At TUM, this project has received funding from the European Research Council (ERC) under the European Union's Horizon 2020 research and innovation programme (grant agreement No. 864234), from Deutsche Forschungsgemeinschaft (DFG, German Research Foundation) under Germany´s Excellence Strategy – EXC 2089/1 – 390776260 and the Walter Benjamin Programme (HE 8878/1-1), TUM.Solar in the context of the Bavarian Collaborative Research Project Solar Technologies Go Hybrid (SolTech), and from the Federal Ministry of Research, Technology and Space (BMFTR, Germany) within the project "SINATRA:CO2UPLED" (project number 033RC034). F.D. and L.E. acknowledge financial support from the Deutsche Forschungsgemeinschaft (DFG) under the Emmy Noether Program (Project 387651688). T.E. acknowledges funding by the Centre for Molecular Water Science (CMWS) within an Early Science Project. The authors further acknowledge the Gauss Centre for Supercomputing e.V. for funding this project by providing computing time through the John von Neumann Institute for Computing on the GCS Supercomputer JUWELS at Jülich Supercomputing Centre. In addition, we gratefully acknowledge the participation and support of the following individuals during beamtimes: Lars Schneider, Hendrik Brockmann, Garret May, and Matthias Nuber. We acknowledge European XFEL in Schenefeld, Germany, for provision of X-ray free-electron laser beamtime at Scientific Instrument FXE (Femtosecond X-ray Experiments) under proposal numbers 2999 and 4079 and would like to thank the staff for their assistance.

**Author Contributions**

VFK, VS, PS, and PK contributed equally. VFK synthesized and characterized the samples and coordinated the XFEL poly-$BiVO_4$ tr-XRD campaign. VS coordinated the XFEL tr-XAS measurement campaign and analysis. PS performed and interpreted DFT calculations of $BiVO_4$ phase stability, electronic structure, and XAS spectra with support from FSH, FPD, and DAE. PK performed data analysis and provided analysis support during the XFEL beamtimes, performed strain field simulations and analysed the tr-XRD data with support from FD and IDS. VFK, VS, PK, LE, TAZ, MWH, DDMG, GZ, ES, and SB formed the on-site XFEL measurement user team, with beamline operation and data analysis support by YJ, YU, TE, FL, XH, DV, FAL, and PZ. BG led synchrotron tr-XRD measurements, with on-site participation by VFK. BG performed Rietveld refinements. DS, HI, and LE conducted lab-based pump–probe measurements in preparation for the beamtime. VS, FD, and PK prepared the figures for the manuscript. FD led the development of XFEL beamtime proposals. FD and IDS developed the experimental concept and specification of experiment parameters for the XFEL beamtimes. DAE, FD, and IDS jointly supervised the project. All authors contributed to preparation of the manuscript. FD and IDS conceived the idea for the project.

**Competing Interests**

The authors declare no competing interests.